# Using Auxiliary Data to Guide the Recruitment of Sites for Randomized Controlled Trials

Robert B. Olsen
(The George Washington Institute of Public Policy, George Washington University)

Maria L. Vasquez-Rossi
(The Trachtenberg School of Public Policy and Public Administration, George Washington University)

**Abstract**

Sampling methods such as stratified random sampling (SRS) can be used to select representative samples of schools for randomized controlled trials (RCTs) of educational interventions. However, these methods may still yield external validity bias when participation by schools is voluntary and participation decisions are associated with unobserved variables. This paper offers a new sampling method called Stratified Random Sampling with Quotas (SRSQ). Under SRSQ, quotas are set to avoid including "too many" schools of a particular type, as defined by auxiliary variables that are unobserved in the sampling frame, but whose population distribution can be estimated from external data. Our simulations find that when the auxiliary variables affect whether or not a school participates in the study, quotas set based on those variables reduce external validity bias. These results suggest that when auxiliary data are available on strong impact moderators for an RCT's target population, these data can be used to address non-ignorable self-selection by schools into RCTs.

Acknowledgments: The authors would like to thank Dan Litwok of Abt Associates for providing the data used in the analysis and The Trachtenberg School of Public Policy and Public Administration at George Washington University for supporting this research.

**Introduction**

Surveys often use "auxiliary data" —data assembled before the survey is conducted—to minimize the bias and variance of the survey estimates. These data may be used: (1) at the sampling stage to divide the sampling frame into strata and sample proportionately from each stratum; (2) during survey operations to prioritize cases for completion before the survey fielding period ends; and (3) after data have been collected to support statistical adjustments for survey nonresponse. For a review of different uses of auxiliary data in surveys, see Cornesse (2020). More generally, auxiliary data are used to obtain more representative survey samples and more accurate survey estimates for the population of interest.

Just as surveys usually suffer from survey nonresponse—a problem on the rise in survey research— randomized controlled trials (RCTs) usually suffer from high rates of noncooperation by study sites that the research team would like to include. This scenario is especially problematic if the factors that influence district and school decisions about whether to participate are also strong moderators of the intervention's impacts. In these cases, districts and schools with especially large or small impacts will be overrepresented in the sample.

Education researchers often begin by recruiting school districts to participate, but district cooperation rates tend to be low.[1] Participation in RCTs can involve extensive commitments that districts are often reluctant to make. Alternatively, districts may be uninterested in testing the intervention that will be subjected to a randomized trial, or they may have objections to random assignment itself. Furthermore, the factors that lead some districts and schools to participate and other to decline may be related to the impacts of the intervention, and thus "nonignorable" in the statistical sense. Thus, noncooperation by districts and schools can lead to randomized trials of educational interventions that poorly represent the populations of students and schools that motivate these studies (Stuart et al., 2017; Tipton et al., 2021).

---

[1] For example, in a recent RCT of academic language interventions, 44 of the 50 districts recruited to participate declined to participate or never responded to outreach by the study team (Corrin et al., 2022, see Appendix, Exhibit B.1). Similarly, in a recent RCT of teacher coaching based on classroom videos, of the 345 districts recruited to participate, only 28 expressed interest in participating: 142 never responded to the outreach and 175 declined to participate (Clark et al., 2022, Appendix, Exhibit B.1).

Recent guidance from a report issued by the Institute of Education Sciences is designed to help researchers address these challenges (Tipton & Olsen, 2022). For example, the report recommends using the Common Core of Data (CCD) from the National Center for Education Statistics (NCES) and other population datasets to build sampling frames, create sampling strata, sample within strata, and construct weights. These datasets can be used to minimize and adjust for differences in characteristics between the sample and population. Minimizing these differences will reduce external validity bias when these characteristics affect both (1) decisions about whether to participate in the study and (2) the impact of the intervention.

However, stratification can mitigate the problem only for variables included in the sampling frame. For variables that are available for all sites in the population, researchers can construct strata and sample proportionately within strata: This will ensure that the sample is similar to the population on these variables. However, in most studies, some and maybe many of variables expected to influence both site participation decisions and the impact of the intervention are not available to include in the sampling frame, and thus cannot be used to stratify the sample. In education RCTs, these could include characteristics of the schools and communities that are not in the CCD and other population datasets and characteristics of the "counterfactual condition" as reflected in the collection of programs and interventions already in place in the districts and schools that might participate in the study. For these variables, stratified random sampling will not necessarily yield samples that are similar to the population.

The available data for building sampling frames typically lack the information that would be most useful to the generalizability of randomized trials: data on strong moderators. This is hardly surprising since general purpose datasets were not developed with that objective in mind. Without data on strong moderators, efforts to ensure the sample resembles the population are necessarily restricted to weak moderators, and good balance between the sample and population on weak moderators does not ensure that the average impact in the sample—the Sample Average Treatment Effect (SATE)—will be close to the average impact in the population, the Population Average Treatment Effect (PATE). In fact, balancing the sample with the population on weak moderator may increase the variance of the impact estimates without substantially reducing the external validity bias.

**Stratified Random Sampling with Quotas**

This paper develops and tests a new method for using auxiliary data to guide site recruitment in randomized trials. The method, which we call Stratified Random Sampling with Quotas (SRSQ), augments Stratified Random Sampling (SRS) in situations where the sampling frame includes all members of the study's target population but only weak moderators of the intervention's impact. SRSQ uses auxiliary data on presumptively stronger impact moderators to calculate and apply quotas for the maximum number of sites of different types, defined based on the values of the auxiliary variables, to include in the study. This method is not likely to substantially improve on SRS when all schools are required to participate. However, SRSQ will likely to improve on SRS when participation by schools in an RCT is voluntary and the participation rate is lower for schools with certain values of strong moderators than others. In these scenarios, SRSQ is designed to ensure that schools with different possible values of the auxiliary variables are represented in the sample in proportion to their representation in the population.

If data on strong moderators were available to include in the population frame, SRSQ would be unnecessary: these moderators could be used to stratify the population and sample from within strata. SRSQ is designed for situations where *auxiliary data contains presumptively strong moderators that cannot be incorporated into the sampling frame,* either due to data restrictions (e.g., based on confidentiality concerns or the proprietary nature of the data) or because the auxiliary data come from a representative sample from the full population of schools that are eligible for the impact study, and thus are not available for every school in the population. In education, promising auxiliary data sources include administrative data maintained by state departments of education and data from surveys conducted by NCES. Quotas may be calculated from the raw data, when the data are made available to outside researchers, or from published reports, when they include sufficient information on the distribution of strong moderators of the intervention's impact.

*Under SRSQ, quotas are set by calculating the share of schools in the population with different values of each moderator variable based on auxiliary data and multiplying these shares by the target sample size.* Under SRSQ, these quotas would be used to exclude some schools that have been sampled and recruited but not yet included in the study sample—those with characteristics for which one or more of the sample quotas have already been met.

Schools would be selected, contacted by field staff, and screened for eligibility as they normally would to confirm that they meet the study's eligibility criteria. Then, field staff would ask these schools a small number of supplemental questions needed to classify them into different categories based on presumptively strong moderators. Eligible schools would be excluded from the study if including them would exceed one or more of the quotas.

To illustrate how SRSQ would work, suppose that researchers were planning a randomized trial in 100 elementary schools to estimate the effects of a new math technology intervention; and the impact of the intervention is expected to depend in part on whether the school has a dedicated technology specialist in the building. The RCT that aims to include 100 schools that on average resemble the population of all eligible schools on total enrollment, the percent of students eligible for free-or-reduced-price meals, and the percentage of schools that have a technology specialist in the building.

The research team could stratify on the first two variables since they are available in the CCD, but not on the third since it is not publicly available from any source for all eligible schools. Schools in the population could be divided, for example, into four equal-sized strata: (1) above median enrollment and above median percent FRPL, (2) above median enrollment and below median percent FRPL, (3) below median enrollment and above median percent FRPL, (4) below median enrollment and below median percent FRPL. To obtain a sample that represents the population on the two stratifying variables, the researchers could aim to include 25 schools from each stratum in the sample. In addition, suppose that a survey conducted by NCES indicates that 66 percent of all schools in the RCT's target population have a technology specialist. To obtain a sample that mirrors the proportion of schools with and without a math technology specialist in the population, the research team could aim for a sample in which 66 out of the 100 schools have a technology specialist. To support this aim, the study team could set quotas of 66 schools with a technology specialist and 34 schools without such a specialist. Finally, under SRSQ, and with these strata and quotas, the research team would:

- Recruit schools in random order within each stratum simultaneously up to the study's targets of 25 schools per stratum;
- Ask each recruited school whether they have a technology specialist:
    - If the school has a technology specialist, include the school if it is willing to participate *and the study has not yet reached its quota of 66 schools with a technology specialist*;
    - If the school does not have a technology specialist, include the school if it is willing to participate *and the study has not yet reached its quota of 34 schools without a technology specialist*.

**Contribution of the Paper**

The primary question addressed by this paper is whether quotas can reduce the external validity bias that would otherwise result from the process by which schools opt in or out of participation in RCTs when these decisions are influenced by strong moderators. While this research is motivated by the challenge of obtaining representative samples for randomized trials, the analysis conducted for this paper is relevant to any type of study that aims to obtain a representative sample on a collection of variables, some of which are available in the study's sample frame while others are not, and where participation in the study by schools is voluntary.

To address this question, we conduct simulations to compare SRSQ to SRS. This comparison holds fixed the stratification of the sampling frame, the sample size targets for each stratum, and the method of selecting schools for recruitment within each stratum (random). The only difference between SRSQ and SRS is whether auxiliary data on presumed impact moderators are used to set and impose quotas to exclude schools that were selected randomly, confirmed to be eligible, but found to have a characteristic for which one or more of the study's quotas have already been met.

For both SRS and SRSQ, we estimate external validity bias, variance, and mean squared error (MSE) for the population mean of three types of variables: (1) auxiliary variables used to set quotas, (2) sampling frame variables uses to construct strata, and (3) variables that are not observed by the researchers. While SRSQ was designed to reduce external validity bias in auxiliary variables, it would be important to also learn whether it reduces or increases the bias in other variables that may moderate the impact of interventions. Also, while external validity bias is our primary performance measure, we also estimate the variance of the sample means generated by the two sampling methods to learn if the likely reductions in bias, at least for auxiliary variables, are accompanied by increases in variance. If so, the MSE may help to mediate whether SRS or SRSQ should be preferred.

While RCTs should be most interested in the bias of the intervention's estimated effects on key outcomes, we focus on the bias in the sample means because the data used in these simulations does not contain the impact of any known intervention, and because the most we can expect from any sampling method is to produce a sample that is similar to the population in its baseline characteristics. However, if SRSQ produces samples that are more similar to the population than SRS on presumptively strong moderators, then we should expect it to also reduce the external validity bias in impact estimates for any intervention for which those presumptively strong moderators are in fact strong moderators.

Finally, the paper also measures the number of schools recruited to achieve a particular sample size. By excluding schools that exceed one or more quotas, SRSQ may increase the number of schools recruited, recruitment costs, and the overall costs of conducting the study. We estimate the number of additional schools recruited under SRSQ so researchers can make informed decisions about whether to adopt it, given the likely cost implications.

**Data**

To compare the performance of SRSQ to SRS, we used data from an extract from the 2018-19 Common Core of Data (CCD), School Universe Survey. This extract, which was constructed and used in prior analyses of sampling methods, included all 42,752 K-5 public schools that were operating in 2018-19 (Litwok et al., 2022).

From these data, we used three variables: (1) total school enrollment, (2) district expenditures per pupil, and (3) percent of the school's students who are eligible for free- or reduced-price meals (hereafter referred to as percent FRPL). Because all three of these variables are in in the CCD, all three could be used to stratify the population in selecting samples for RCTs. However, to compare the performance of SRSQ to SRS, *we pretended that only one of them was part of the sampling frame and available to use in defining strata*. Of the other two variables, one was treated as an auxiliary variable to be used in setting quotas and the other was treated as unobserved and thus unavailable for use in defining strata or setting quotas.

**Methods**

Using the data described above, we applied simulation methods to compare the performance of SRS and SRSQ. To compare their performance, we calculated three performance measures:

1. **External validity bias**, defined as the expected difference between (1) the sample mean and (2) the population mean.[2]

2. **Variance**, defined as the expected squared difference between (1) the sample mean and (2) the expected value of the sample mean.

3. **Mean squared error**, defined as the expected squared difference between (1) the sample mean and (2) the population mean.

These performance measures were calculated separately for each combination of:

- **Sampling method (SRS or SRSQ).** This allowed us to directly compare the performance of the two methods for different variables and populations, as described below.

- **Variable type (stratifier, auxiliary variable, or unobserved variable).** This allowed us to test whether SRSQ performed better than SRS for the variable used to construct quotas, as we expected; it also allowed us to test whether SRSQ performed better or worse than SRS for (1) the stratifying variable and (2) the unobserved variable used neither in constructing strata nor in setting quotas.

- **Population.** Samples were drawn from 51 different populations of K-5 schools—(1) the entire U.S., including all 50 states and the Bureau of Indian Affairs, (2) each of the 50 states except Hawaii,[3] and (3) the District of Columbia (DC). Comparing the methods with different populations allowed us to assess whether the relative performance of the two sampling methods was fairly consistent across populations, or whether the relative performance varied across populations—ideally in ways that would allow us to glean the conditions under SRSQ performs best and thus should be used in practice.

---

[2] Since the variables in each population were standardized to have a mean of zero and standard deviation of 1, the external validity bias simplifies to the expected value of the sample mean.

For the simulations, we needed to decide which of the three variables described earlier to use in stratifying the population, which variable to use in setting quotas, and which to use for neither. To ensure that our simulation results were not driving by this arbitrary decision, we ran simulations for SRS and SRSQ for each of the six possible ways of assigning the three variables to the three roles and averaged the performance measures across these six pairs of simulations (two different sampling methods by six variable combinations). For each sampling method, we calculated the bias and variance for each of the six variable combinations, took the simple average across those six estimates, and report those estimates in this paper. estimates reported in this paper were calculated as the simple average of those measures across the six simulations, one for each variable co for the six sets of simulations.

---

[3] We excluded Hawaii from the state-level populations because all schools in Hawaii belong to the same school district, so the variable expenditure per pupil, defined at the district-level did not vary across schools.

**Table 1.** Simulation design for comparing SRSQ to SRS.

| Simulation Pair | Method | Variables Used in Simulations | | |
|---|---|---|---|---|
| | | Observed Variable Used to Construct Strata | Auxiliary Variable Used to Set Quotas | Unobserved Variable Not Used in Simulations |
| 1 | SRS | Percent FRPL | Total enrollment | Expenditures per pupil |
| | SRSQ | | | |
| 2 | SRS | Percent FRPL | Expenditures per pupil | Total enrollment |
| | SRSQ | | | |
| 3 | SRS | Total enrollment | Percent FRPL | Expenditures per pupil |
| | SRSQ | | | |
| 4 | SRS | Total enrollment | Expenditures per pupil | Percent FRPL |
| | SRSQ | | | |
| 5 | SRS | Expenditures per pupil | Percent FRPL | Total enrollment |
| | SRSQ | | | |
| 6 | SRS | Expenditures per pupil | Total enrollment | Percent FRPL |
| | SRSQ | | | |

To account for natural variation across possible samples, we selected 1,000 SRS samples and 1,000 SRSQ samples for each of 306 combinations of simulation parameters (51 populations times 6 allocations of variables to roles in the sampling process). This comes to 612,000 total samples selected for the study. Earlier analyses were conducted with 500 SRS samples and 500 SRSQ samples. The external validity bias estimates were very similar for 500 samples as they were for 1,000 samples, suggesting that 1,000 samples are sufficient to obtain stable estimates.[4]

An important decision in the design of these simulations is the target sample size of each sample. We set the target sample size to 100 schools, which is toward the high end of the range for RCT samples in education. In small studies, researchers are most likely to rely entirely on existing connections to districts and schools to recruit a sample. Furthermore, small samples chosen randomly may differ substantially from the population from which they were selected simply due to random chance. For these reasons, we focus on the relative performance of SRS and SRSQ in large samples where random sampling is most likely to be considered. We refer to 100 as the "target" sample size because in some cases, the target might not be reached due to nonparticipation by schools and the exclusion of schools that would exceed one of the quotas.

---

[4] For a comparison of the external validity bias and other metrics between 500 samples and 1,000 samples, see the appendix.

To test the relative performance of SRSQ versus SRS when sites make nonignorable decisions about whether to participate, we set the probability of agreeing to participate in the study to 0.5 for schools in the bottom half of schools in the population based on the auxiliary variable and 0.25 probability for schools in the top half of schools in the population based on this variable. These probabilities were set in this way to bias the sample mean for that variable downward.[5]

To select a sample using SRS, we:

- **Stratified the population.** In particular, we divided the population of N schools into five quintiles containing approximately N/5 schools, based on the stratifying variable. Then weset stratum-specific sample size targets that were proportional to the share of schools in the population within each stratum. This yielded sample size targets of approximately 20 schools per stratum.

---

[5] Agreement probabilities could have been set twice as high for schools in the top half of the distribution than for schools in the bottom half of the distribution to bias the sample mean of the auxiliary variable upward. We would expect these probabilities to yield results to those reported here in terms of the absolute value of the external validity bias.

- **Ranked schools in random order within strata for recruitment.** We placed schools in random order, assigning the first school each stratum a rank of 1 and the last school a rank of $N_s$, where $N_s$ equals the total number of schools in stratum s.

- **Pooled schools across strata and created a single ordered list of schools.** This ordered list determined the sequence in which schools were recruited to participate in the study. Schools were ordered first by within-stratum rank (see previous step). This step created N/5 groups of five schools with the same rank—one for each stratum—starting with schools with a rank of 1. This ensured that for each of the five strata, schools were recruited in random order, since schools were ranked in random order within strata (see previous step). Then, among schools with the same rank, we sorted the schools randomly. This[6] was designed to mimic a recruitment process in which schools from each stratum were being recruited simultaneously (instead of prioritizing schools from particular strata).[7]

- **Recruited schools that were willing to participate until sampling was completed.** Schools were recruited one at a time in the order placed in the previous step. To simulate participation decisions, each school was assigned a uniform random number between 0 and 1. A school agreed to participate if and only if the random number was below the probability of agreement. The process of selecting schools and simulating their participation decisions continued until the sample size target for each stratum was met or until all schools in the stratum had already agreed or declined to participate.

---

[7] This latter step was designed to avoid situations where quotas were satisfied in strata recruited early in the process, limiting the inclusion of schools from strata recruited later in the process.

Table 2 shows a hypothetical example with 20 schools in the population. In this example, the 20 schools are grouped into five quintiles/strata and ranked in random order (left-most column). Then they are pooled across strata, sorted by their within-stratum rank, and then sorted randomly among schools with the same within-stratum rank (right column) to create a single ordered list for recruitment.

**Table 2.** Illustration of how schools are rank ordered for recruitment.

| Rank Ordering of Schools Within Strata (school x.y is the y[th] school in stratum x) | Rank Ordering of Schools Across Strata (ordered by rank and randomly within rank) |
|---|---|
| **Stratum 1 Schools** | School 2.1 ⎤ |
| School 1.1 | School 4.1 ⎟ |
| School 1.2 | School 1.1 ⎬ **Rank 1 Schools** |
| School 1.3 | School 5.1 ⎟ |
| School 1.4 | School 3.1 ⎦ |
| **Stratum 2 Schools** | School 3.2 ⎤ |
| School 2.1 | School 1.2 ⎟ |
| School 2.2 | School 4.2 ⎬ **Rank 2 Schools** |
| School 2.3 | School 2.2 ⎟ |
| School 2.4 | School 5.2 ⎦ |
| **Stratum 3 Schools** | School 3.3 ⎤ |
| School 3.1 | School 1.3 ⎟ |
| School 3.2 | School 5.3 ⎬ **Rank 3 Schools** |
| School 3.3 | School 2.3 ⎟ |
| School 3.4 | School 4.3 ⎦ |
| **Stratum 4 Schools** | School 1.4 ⎤ |
| School 4.1 | School 3.4 ⎟ |
| School 4.2 | School 5.4 ⎬ **Rank 4 Schools** |
| School 4.3 | School 4.4 ⎟ |
| School 4.4 | School 2.4 ⎦ |
| **Stratum 5 Schools** | |
| School 5.1 | |
| School 5.2 | |
| School 5.3 | |
| School 5.4 | |

To select a sample using SRSQ, we ranked strata and schools within strata in the same random order as for SRS, and we recruited schools in the same order described above and illustrated in Table 2. However, with SRSQ, we also:

1. **Estimated the population distribution of the auxiliary variable.** For the auxiliary variable, we estimated the thresholds between five quintiles in the population. By construction, these thresholds divided the population into five groups with an equal number of schools—or as close to equal numbers as possible given the distribution of the quota variable.[8]

2. **Set quotas based on the population distribution.** Specifically, we set quotas of approximately 20 schools per quintile to match the distribution of schools in the population and align with the overall target sample size of 100 schools.

3. **Apply the quotas during recruitment.** In selecting schools in sequence within a stratum, a school was skipped and excluded from the sample—even if it would be willing to participate—if it falls within a quintile for the auxiliary variable for which the quota of ~20 schools per quintile had already been reached.

Returning to the hypothetical example from Table 1, we could imagine a scenario under which School 2.4 in Stratum 2 would be willing to participate and included under SRS but excluded under SRSQ. This could occur if, for example, quotas were set based on total enrollment in the school, School 2.4 falls into the quintile of the largest schools based on total enrollment, and the quota for the largest schools has already been met—so School 2.4 must be excluded to enforce the quotas.

---

[8] The quintiles are more likely to be more unequal in size within small states where the number of school districts was relatively small or in states where the values of the variables have uneven distributions, for instance a state with a large proportion of schools where 100% of students are eligible for free/reduced price lunch.

**Simulation Results**

We first present the simulation findings for the samples selected from the all K-5 schools in the U.S., to inform large national studies. Then we present the findings separately by state. The state-level findings directly inform studies that aim to produce estimates for state-wide populations. But because the number of schools in the population vary enormously across states, the state-level findings provide insights on how the size of the population affects the relative performance of SRS and SRSQ.

**Findings from national population.** For the U.S. as a whole, we find that quotas dramatically reduce the magnitude of the external validity bias in the auxiliary variable (Table 3). Recall that the auxiliary variable is the presumptively strong moderator that was used to vary the school agreement probabilities. As a result, SRS generated samples that differ substantially from the population in the mean value of the auxiliary variable, as reflected in the external validity bias for SRS of 0.2386 standard deviations. In contrast, by setting quota for the number of schools that can be included for each population quintile of the auxiliary variable distribution, SRSQ reduced the external validity bias to near zero (0.0064 standard deviations). These results suggest that quotas achieved their primary objective of reducing external validity bias in the auxiliary variable.

Furthermore, for the national population, quotas also reduced the variance and MSE in the auxiliary variable. As we can see by comparing SRSQ to SRS, quotas reduced the variance in this variable from 0.0083 standard deviations to 0.0023 standard deviations and the MSE from

**Table 3.** Mean performance of sampling methods for U.S. population across 1,000 samples (standard deviation or effect size units).

| Performance Measure and Variable Type | Stratified Random Sampling (SRS) | Stratified Random Sampling with Quotas (SRSQ) | Difference (SRSQ-SRS) |
|---|---|---|---|
| **External Validity Bias (Absolute Value)** | | | |
| Auxiliary Variable (Quota) | 0.2386 | 0.0064 | -0.2322 |
| Sampling Frame Variable (Strata) | 0.0016 | 0.0006 | -0.0009 |
| Unobserved Variable (Neither) | 0.0244 | 0.0018 | -0.0226 |
| | | | |
| **Variance** | | | |
| Auxiliary Variable (Quota) | 0.0083 | 0.0023 | -0.0060 |
| Sampling Frame Variable (Strata) | 0.0023 | 0.0023 | -0.0001 |
| Unobserved Variable (Neither) | 0.0102 | 0.0097 | -0.0005 |
| | | | |
| **Mean Squared Error (MSE)** | | | |
| Auxiliary Variable (Quota) | 0.0652 | 0.0024 | -0.0629 |
| Sampling Frame Variable (Strata) | 0.0023 | 0.0023 | -0.0001 |
| Unobserved Variable (Neither) | 0.0108 | 0.0097 | -0.0011 |
| | | | |
| **Total Number of Schools** | | | |
| Contacted about participating | 267.1 | 454.9 | 187.8 |
| Excluded due to quotas | 0.0 | 160.8 | 160.8 |
| Invited to participate | 267.1 | 294.1 | 27.0 |
| Declined to participate | 167.1 | 194.1 | 27.0 |
| Agreed to participate | 100.0 | 100.0 | 0.0 |

0.0652 standard deviations to 0.0024 standard deviations. The reduction in MSE is primarily due to the large reduction in external validity bias but reinforced by the smaller reduction in variance.[9]

---

[9] For the auxiliary variable, the MSE reduction equals 0.0603 and the variance reduction equals 0.0052. This means that the reduction is variance is responsible for 0.0052 / 0.0603 = 8 percent of the reduction in MSE.

Interestingly, for this population, quotas also somewhat reduced the external validity bias in the variable treated as unobserved and thus not used to define strata or set quotas. For this variable, quotas reduced the bias from 0.0244 standard deviations to 0.0018 standard deviations, a reduction in 0.0226 standard deviations. These estimates reveal an unintended benefit of setting quotas: Quotas that reduce the external validity bias in the auxiliary variable may also reduce the bias in unobserved moderators that are correlated with the auxiliary variable.

For the variable used to construct strata, quotas had little or no effect on bias, variance, or MSE. Quotas reduced the external validity bias for the stratifying variable from 0.0016 standard deviations to 0.0006 standard deviations. But they had no effect on the variance or the MSE in the first four digits beyond the decimal.

The findings described above indicate that in our simulations, quotas substantially improved the performance of random sampling. However, we find that the reduction in external validity bias comes at a substantial cost. Once a school is "contacted" (encountered by program in random order within strata), the value of the auxiliary value is "observed", and the school is excluded if including it would exceed one of the quotas. **On average across the 500 samples, 158.8 schools are excluded due to the quotas.** For each excluded school, a replacement school had to be contacted, confirmed not to exceed one of the quotas, and agree to participate. As a result, the quotas substantially increased the average number of schools that needed to be contacted to achieve the target sample size of 100 schools—from 267.1 with SRS to 454.9 with SRSQ, an increase of 70 percent. In an actual study, a 70-percent increase in the number of schools contacted and vetted would substantially raise the costs of recruiting the sample.

We would anticipate that the additional costs from applying quotas to be proportional in a loose sense to the reduction in external validity bias. If the probability of agreeing to participate is strongly related to the auxiliary variable, the external validity bias will be large, and the number of schools excluded by the application of the quotas will also be large. In this setting, SRSQ will have large benefits in terms of bias reduction, but it will come at a high cost in terms of the number of additional schools contacted and vetted to achieve the target sample size. In contrast, if the probability of agreeing to participate is only weakly related to the auxiliary variable, the external validity bias will be much lower and the number of schools excluded by the application of the quotas—and thus the number of additional schools contacted and vetted to achieve the target sample size—will be much smaller. In this setting, SRSQ will have modest benefits but at lower cost.

**Findings from state populations.** We find that the quotas substantially reduced the magnitude of the external validity bias in the auxiliary variable when sampling from all but the smallest states (Figure 1). For example, in Massachusetts, which had 672 K-5 schools, quotas reduced the absolute bias in the auxiliary variable from 0.235 in SRS to 0.023 for SRSQ. The simulation findings for the largest states (e.g., California and Texas) mirror the findings for the national population for all performance metrics and for all three types of variables. Of particular note are the findings for external validity bias. For all states "larger" than Iowa (with more than 459 eligible schools), quotas reduce the bias in the auxiliary variable from between 0.18 standard deviations and 0.25 standard deviations to less than 0.05 standard deviations.

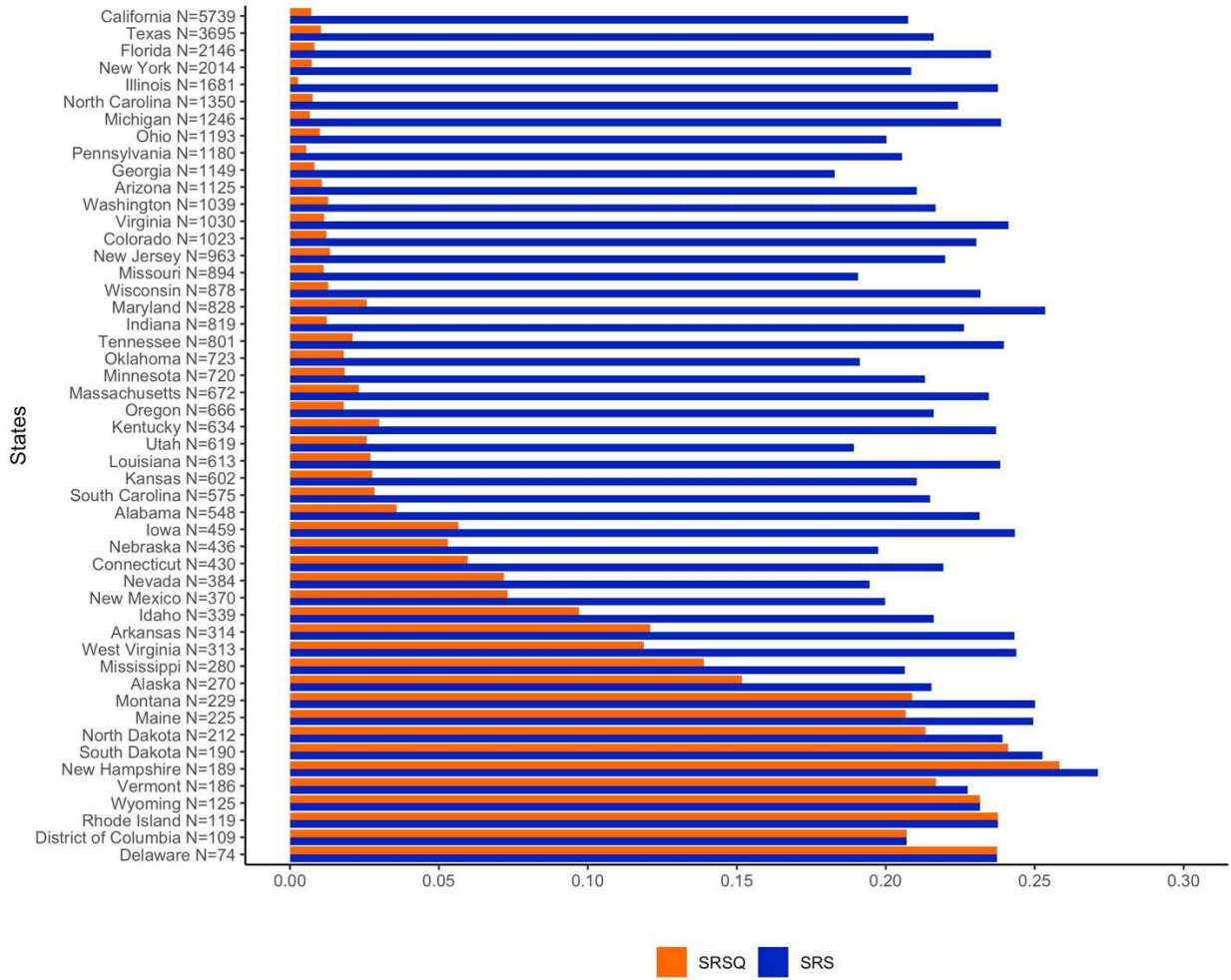

**Figure 1.** External validity bias in the sample mean of the auxiliary variable (absolute value).

In the smallest states, quotas had little effect on the magnitude of the external validity bias in the auxiliary variable because the quotas were almost never binding. When the population is sufficiently small relative to the target sample size of 100 schools, neither SRS nor SRSQ will select enough schools of any type to risk exceeding one of the quotas. For example, in Delaware, the state with the smallest number of K-5 schools (n=74), quotas had little effect on the magnitude of the external validity bias. But Delaware has fewer than 100 K-5 schools in total, so obtaining a sample of 100 schools would be infeasible. South Dakota has 190 K-5

schools—more than 100, so obtaining a sample of 100 schools would be theoretically possible. But since we set the agreement probability to 0.50 for schools above the median on the auxiliary variable and 0.25 for schools below this median, the likelihood of obtaining 100 schools that agree to participate is very low even if all 190 schools were selected. Since the mean agreement probability is 0.375, we focus our attention on states with more than 100/0.375=267 schools so that different sampling methods could yield different samples of schools and different magnitudes of external validity bias.

In states with enough schools such that different sampling methods could plausibly differentiate themselves, quotas substantially reduced the magnitude of the external validity bias in the auxiliary variable. In Alaska—the smallest state with at least 267 schools (n=270)—quotas reduced the magnitude of the external validity bias from 0.215 standard for SRS to 0.152 for SRSQ. The bias reduction is even larger for larger states. For example, in Alabama, which had about twice as many K-12 schools as Alaska (n=548), quotas reduced the magnitude of the external validity bias from 0.232 for SRS to 0.036 for SRSQ. The difference in the effects of the quotas between Alabama and Alaska demonstrates that SRSQ benefits from having a larger population of schools to choose from. In the very largest populations (e.g., the U.S. and each of the three largest states, California, Texas, and Florida), SRSQ yielded external validity bias that was trivially small— 0.01 in absolute value or less.

If the quotas reduced external validity bias but also reduced the attained sample sizes—by excluding schools that were not replaced—and increased the variance of estimates, we would need to weigh the bias-variance tradeoff in deciding whether to prefer SRSQ to SRS. Figure 2 shows that **among the states that are *small but still large enough such that obtaining a*** *sample of 100 schools was plausible*, quotas reduced the achieved sample size by a nontrivial amount.

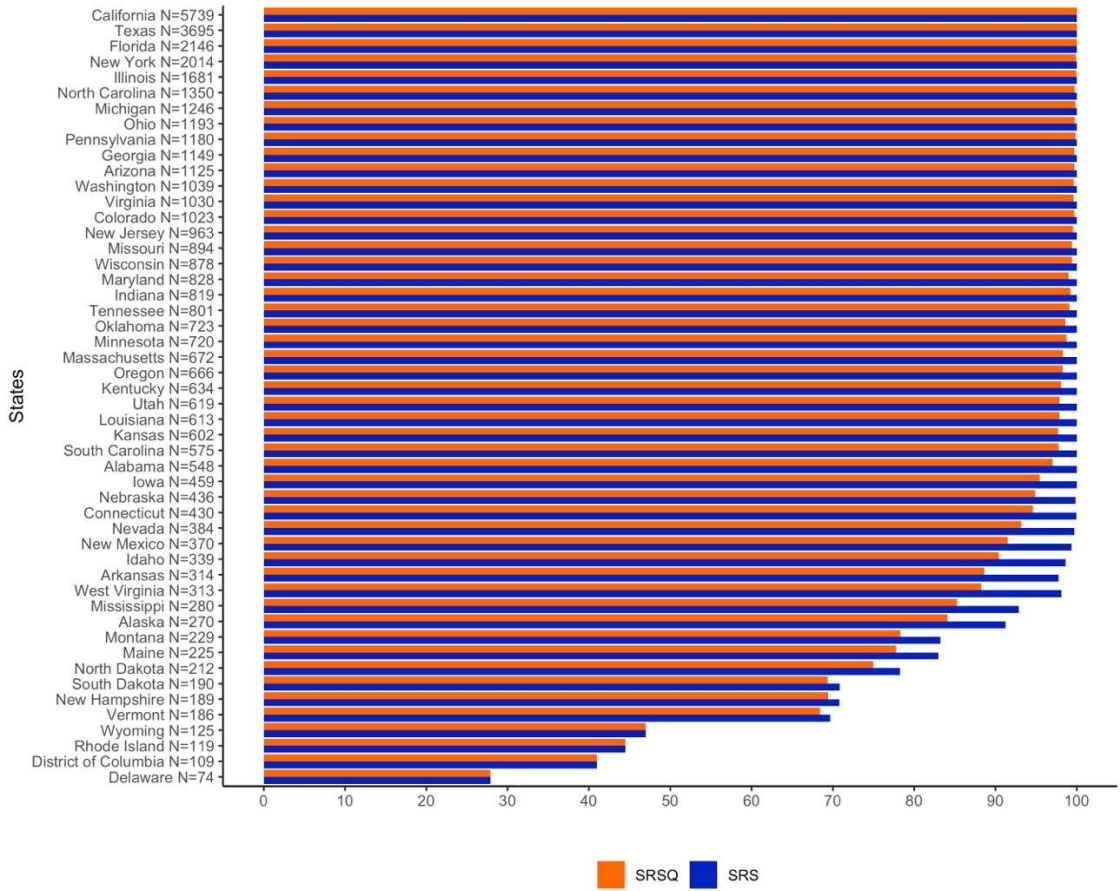

**Figure 2.** Average number of schools included in the sample.

For example, quotas reduced the average sample size in Alaska from 91.24 schools using SRS to 84.12 using SRSQ. However, if we consider Alabama, which has twice as many K-12 schools as Alaska, the reduction in sample size from quota is much smaller, from 100.00 schools to 97.05 schools. For states as large or larger than Kentucky (n=634 K-5 schools), the quotas reduced the achieved sample sizes by fewer than two schools (e.g., from 100 schools to 98.07 schools for Kentucky, with smaller reductions in larger states). These findings show that the reduction in sample size is only notable for a small slice of the size distribution of states. For smaller states, the quotas are never binding, so they do not affect the achieved sample size; for larger states, the quotas are binding, but schools excluded due to the quotas can typically be replaced by other schools in the same stratum, so the effect on the achieved sample size is trivially small.

The big question is whether smaller samples led to less precise estimates of the mean for the auxiliary variable. Figure 3 suggests that in our simulations, the answer was generally no— **even among the states for which quotas reduced the achieved sample size nontrivially, the quotas *reduced* the variance of the sample means for the auxiliary variable**. For example, the quotas reduced the variance of the sample mean of the auxiliary variable across the 1,000 samples from 0.0061 to 0.0054 in Alaska and from 0.0059 to 0.0044 in Mississippi. These findings indicate that in our simulations, the quotas reduced the external validity bias *and* the variance of the estimates. This suggests that the quotas moderated the random differences between the sample and population that can occur with random sampling even in samples with as many as 100 schools.

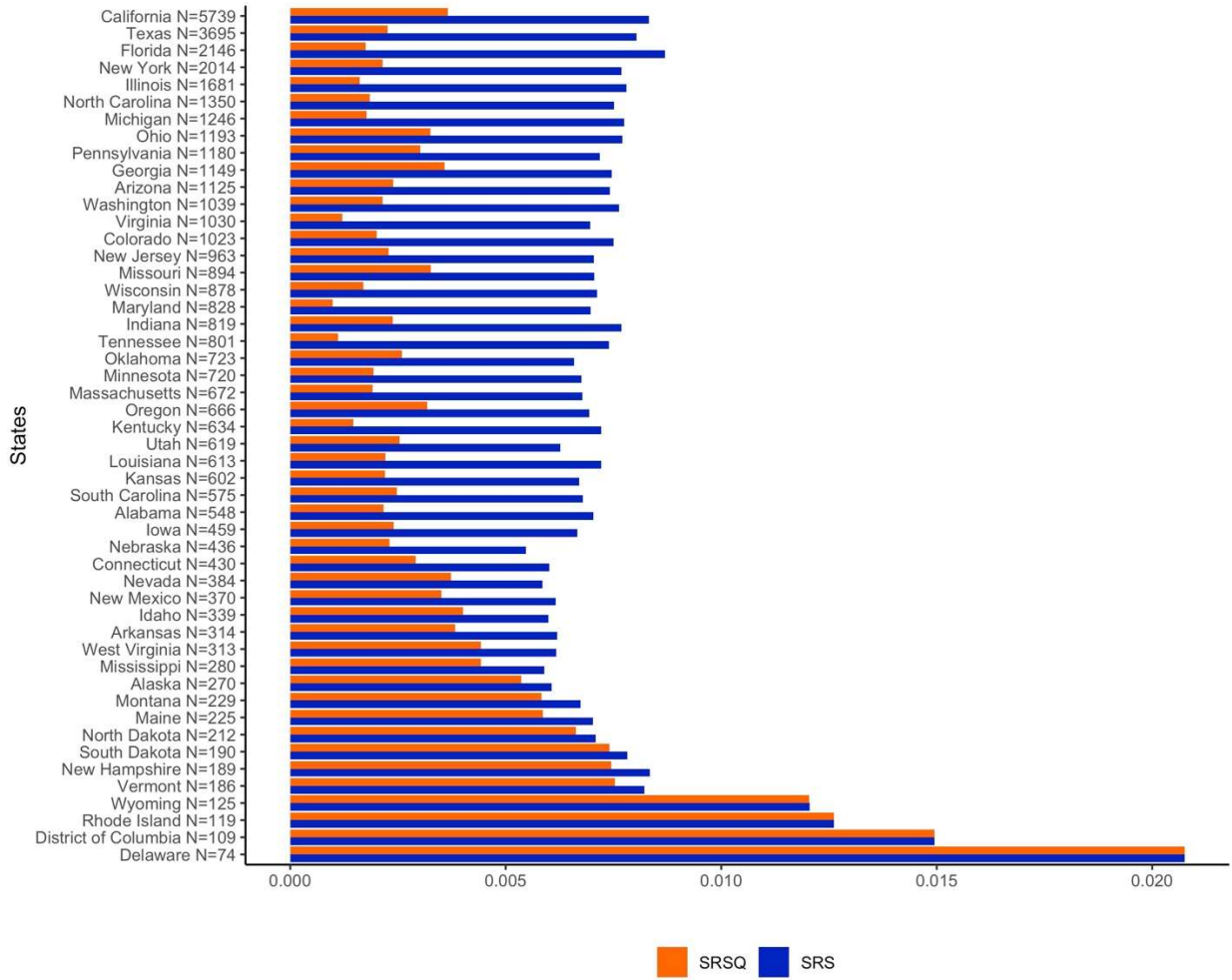

**Figure 3.** Variance of the sample mean of the auxiliary variable.

Lastly, it is important to consider whether setting quotas based on the auxiliary variable affected the external validity bias or variance in estimating the means for *other* variables that were not used to set quotas. Reductions in bias and variance for these variables would reinforce the benefits of quotas reported above; increases in bias or variance for these variables would suggest that quotas can improve the representativeness of the sample with respect to auxiliary variables, but only at the expense of worse representativeness for other variables. In summary, we find that quotas reduce the bias for unobserved variable considerably, with little effect on the bias for the stratifying variables, and only small effects on the variance for either type of variable.

For the stratifier variable, the external validity bias is low for both SRS and SRSQ—so low that the differences are trivial (Figure 4). For the U.S. as a whole (not shown and for large states, the bias for both methods is typically less than 0.01 standard deviations, and typically somewhat smaller for SRSQ than for SRS. For smaller states, the bias for both methods is larger, but less than 0.10 standard deviations, with little difference between the two methods. The differences in the variance of the sample mean for the stratifier variable across the two methods are also trivial (Figure 5): The variance tends to be slightly smaller for SRSQ than for SRS in the large states and slightly larger for SRSQ than for SRS in the small states—states where quotas often prevented the sampling approach from reaching the target sample size.

For the variable treated as unobserved—and used neither for stratification nor for setting quotas—SRSQ led to substantial reductions in external validity bias for the entire U.S. (not shown) and for most states, especially large states (Figure 6). *This suggests that quotas based on auxiliary data can help to obtain samples that better represent the population for unobserved variables that may moderate the intervention's impact*. Presumably, this finding results from the correlation between auxiliary variables and unobserved variables. Relative to SRS, SRSQ yields slightly smaller variances for the sample mean of the unobserved variable for the U.S. as a whole and for larger states (Figure 7); it yields slightly larger variances for some of the smallest states where quotas tended to reduce the achieved sample size.

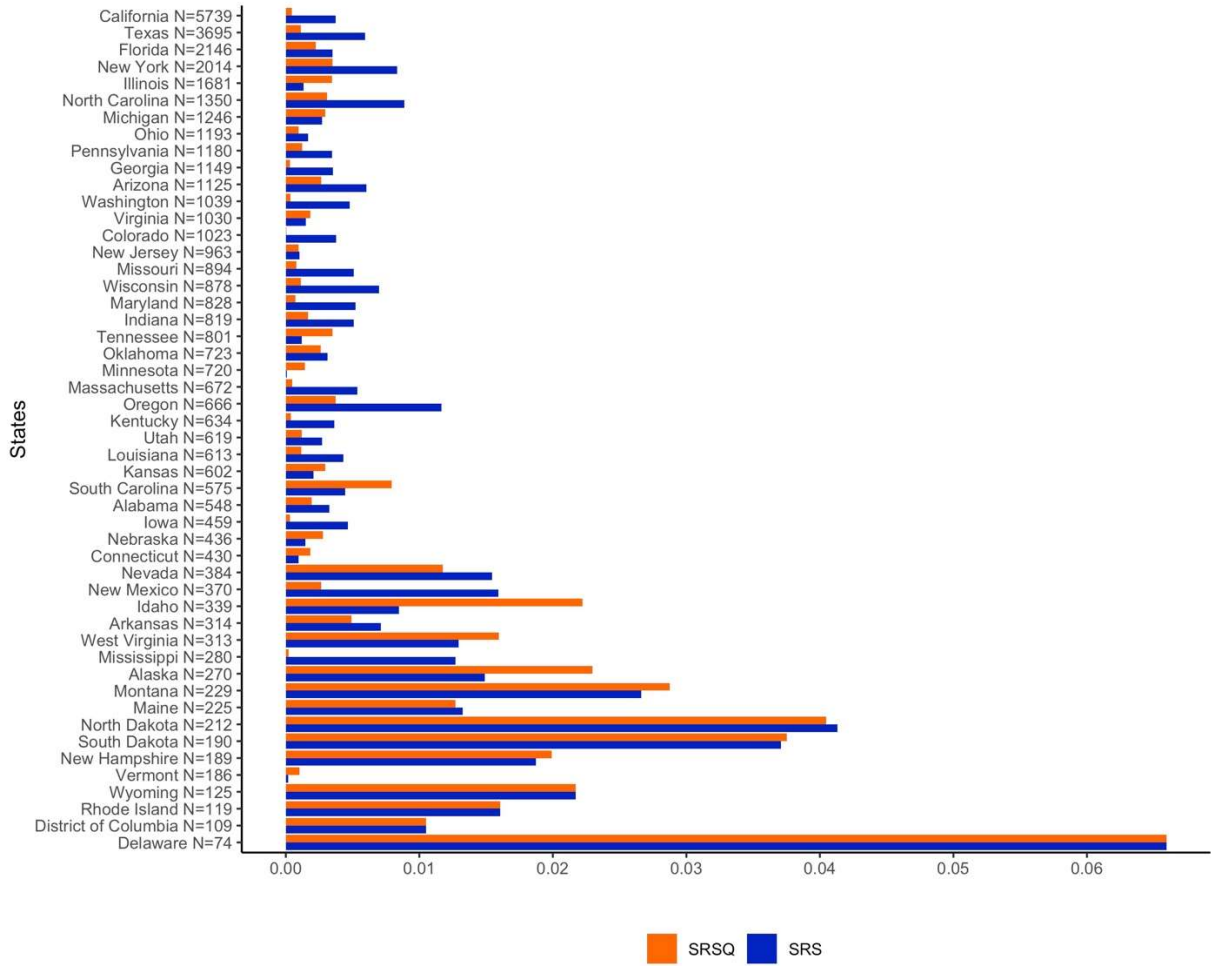

**Figure 4.** External validity bias in the sample mean of the stratifier variable (absolute value).

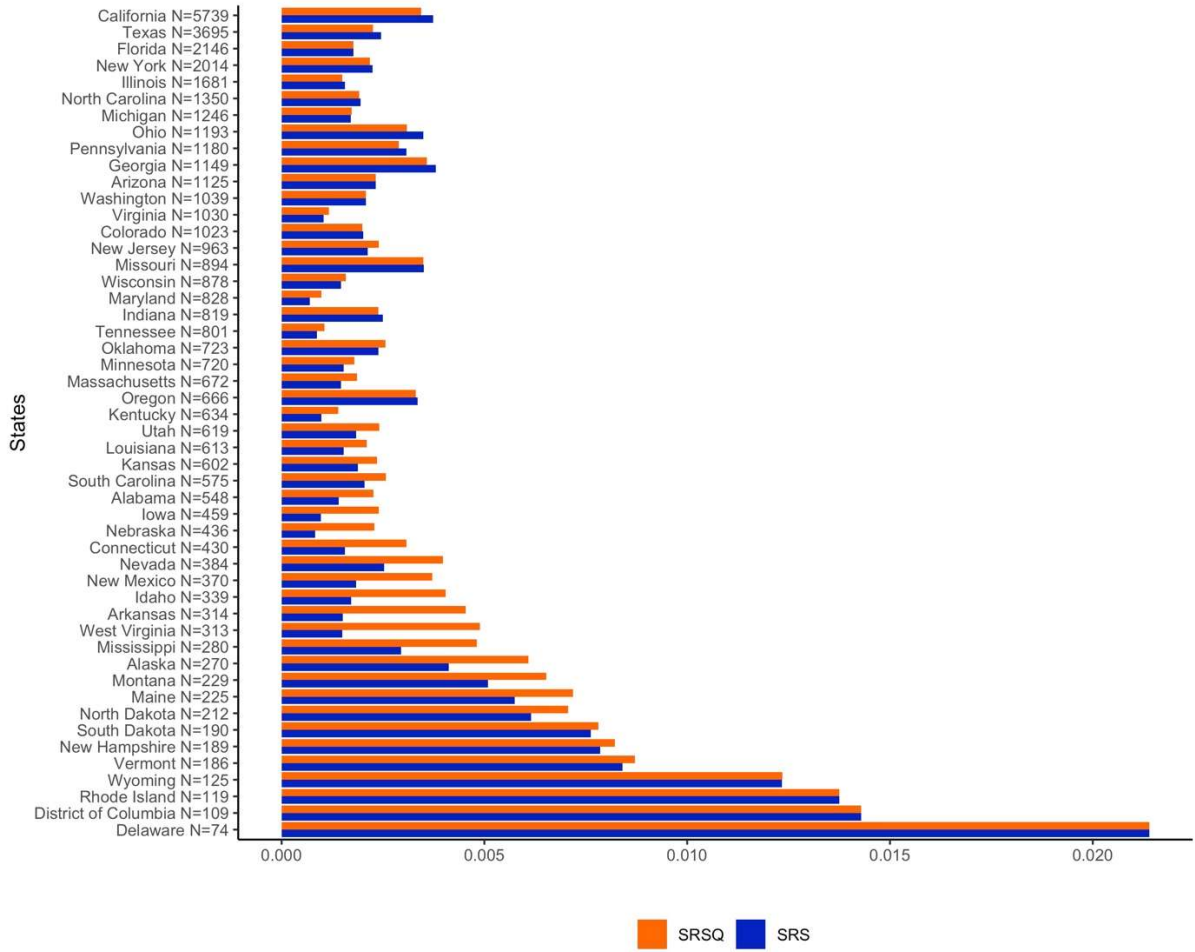

**Figure 5.** Variance of the sample mean of the stratifier variable.

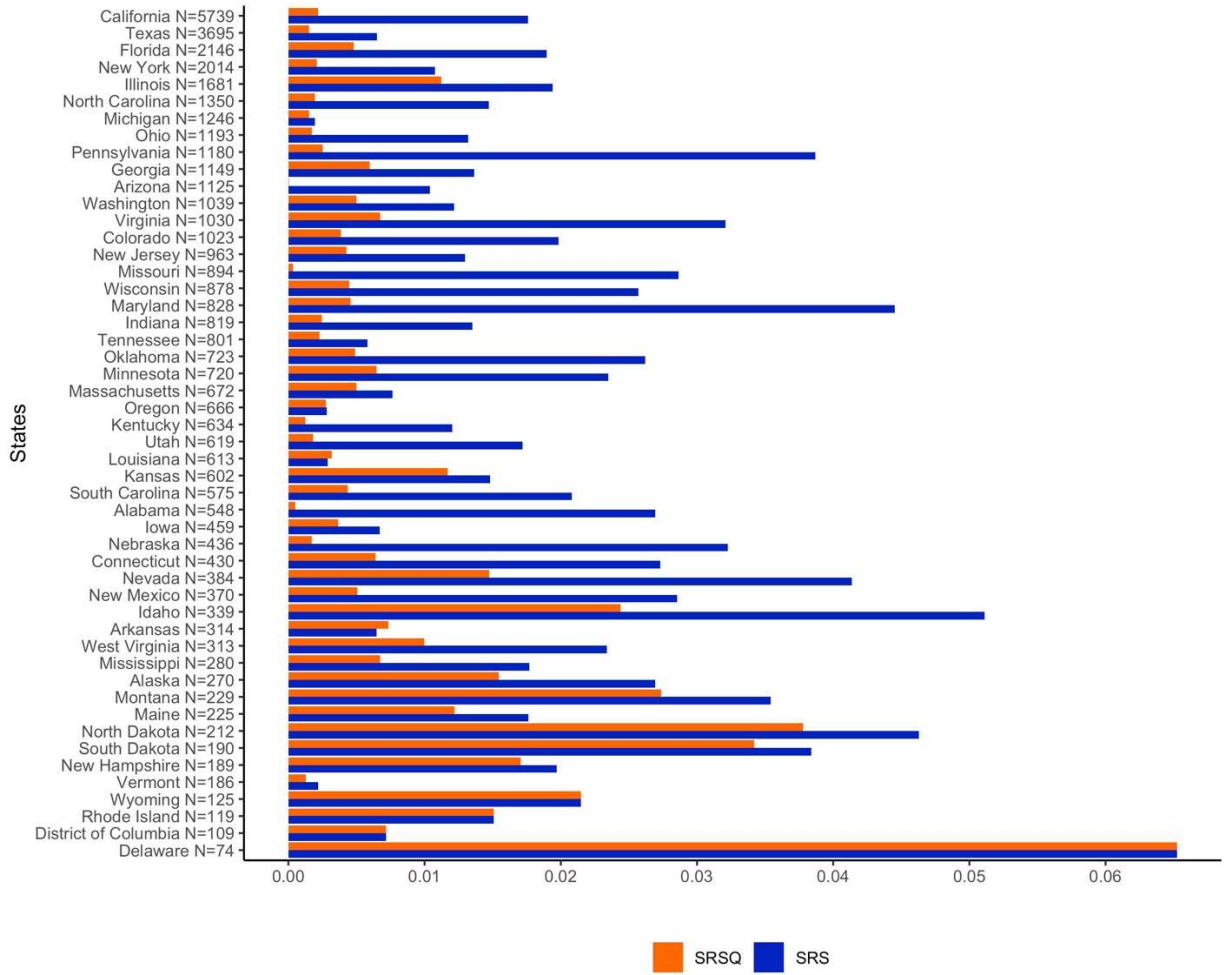

**Figure 6.** External validity bias of the sample mean of the unobserved variable (absolute value).

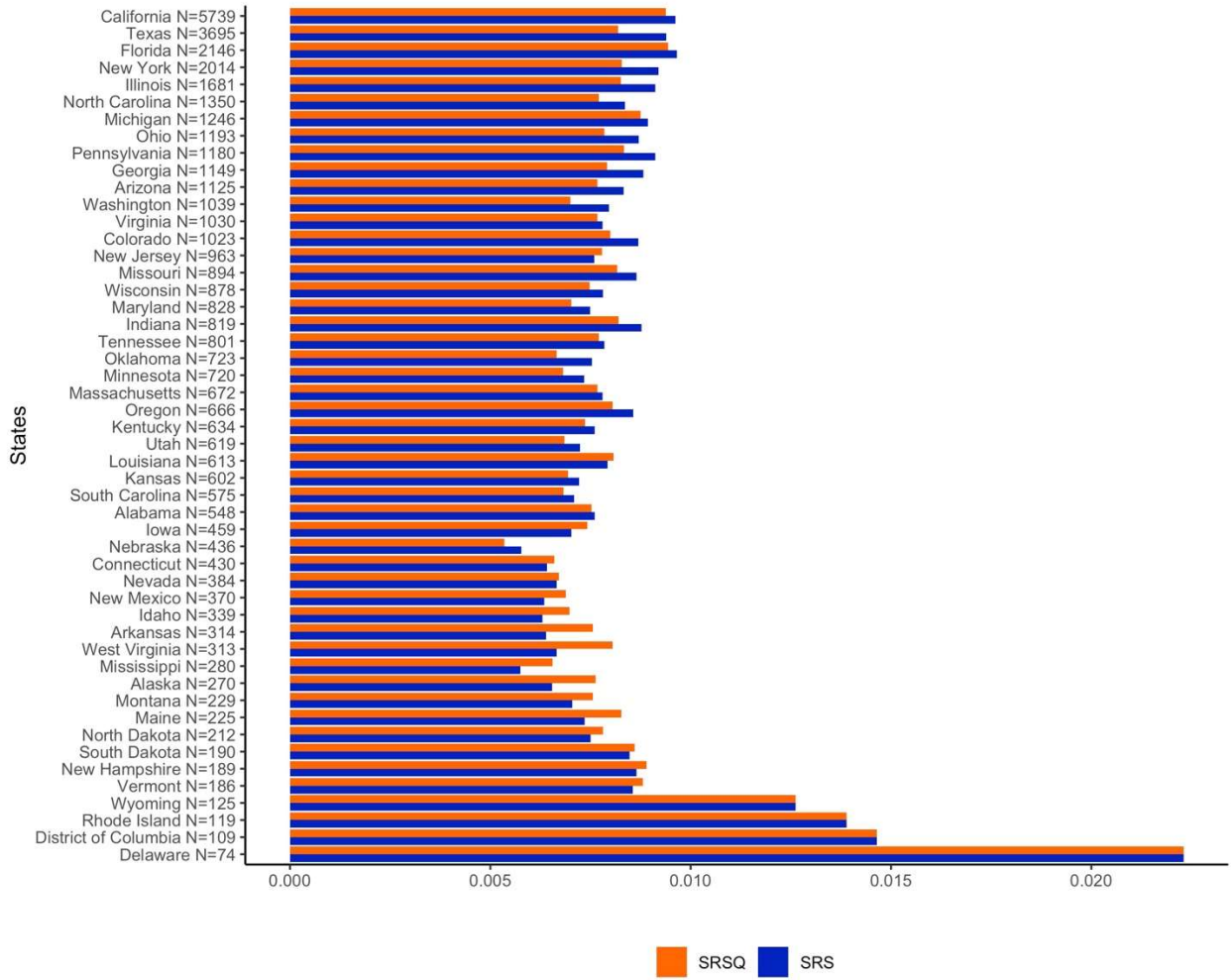

**Figure 7.** Variance of the sample mean of the unobserved variable.

**Discussion**

This paper introduces the idea of augmenting standard sampling methods with quotas to control the composition of the sample. Quotas are set as "not to exceed" caps for certain types of schools based on school characteristics that cannot be observed until a school is selected and contacted about participating, but for which the population distribution is known or can be estimated using external data (e.g., an NCES survey of schools). The findings from this paper indicate that quotas can provide substantial protection against obtaining an unrepresentative sample of the population—both for variables used to set quotas *and* for unobserved factors that are not controlled directly by either stratification or quotas. Furthermore, the paper finds that the reduction in external validity bias for both auxiliary and unobserved variables come without increases in variance. Therefore, the results from this paper suggest that researchers can reduce external validity bias without substantially affecting the precision of their estimates.

The paper does find that applying quotas increases the number of schools that must be recruited to reach the target sample size. This is not surprising, since when the quotas are binding, researchers will have to exclude schools that would be willing to participate, and additional schools must be recruited to take their place. In our simulations, quotas substantially increased the number of schools recruited. The large increases in the number of schools recruited are surely an artifact of the strong relationship constructed between the auxiliary variable and the probability that schools agreed to participate in our simulations. In real world applications, where this relationship is probably weaker, quotas would likely exclude fewer schools than in our simulations, leading to smaller increases in the number of schools recruited to meet the study's sample size requirements (as well as smaller reductions in external validity bias).

While this paper tested quotas in the context of random sampling, quotas may benefit RCTs that do *not* select schools randomly. This feature makes quotas broadly applicable to RCTs that recruit large samples, however they recruit them. Few RCTs in education select random samples of schools; most select schools purposively or for convenience. In these studies, quotas should still protect against obtaining a highly unrepresentative sample due to the factors that lead research teams to recruit some schools instead of others and factors driving schools' decisions about whether to participate. Therefore, the approach described here for setting and applying quotas is applicable in any RCT as long as external information on the distribution of key impact moderators is available.

**Appendix**

The simulations presented in the body of this paper were based on 1,000 simulated samples using SRS and 1,000 simulated samples using SRSQ. We concluded that 1,000 simulated samples of each type would produce sufficient stable results by comparing the relative performance of SRSQ and SRS for 500 simulated samples to their relative performance from 1,000 simulated samples. Table A.1 below shows that these differences are small. Since differences between 1,000 simulated samples and 1,500 simulated samples would be even smaller, we concluded that 1,000 simulated samples were sufficient for our purposes.

Table A.1: Mean Performance of Sampling Methods for U.S. Population across 1,000 Samples (Standard Deviation or Effect Size Units)

|  | Differences between SRSQ and SRS | | Difference (1,000-500) |
|---|---|---|---|
| **Performance Measure and Variable Type** | 500 samples | 1,000 samples | |
| **External Validity Bias (Absolute Value)** | | | |
| Auxiliary Variable (Quota) | -0.2311 | -0.2322 | -0.0011 |
| Sampling Frame Variable (Strata) | -0.0009 | -0.0009 | 0.0000 |
| Unobserved Variable (Neither) | -0.0218 | -0.0226 | -0.0008 |
| | | | |
| **Variance** | | | |
| Auxiliary Variable (Quota) | -0.0061 | -0.0060 | 0.0001 |
| Sampling Frame Variable (Strata) | -0.0001 | -0.0001 | 0.0000 |
| Unobserved Variable (Neither) | -0.0008 | -0.0005 | 0.0003 |
| | | | |
| **Mean Squared Error (MSE)** | | | |
| Auxiliary Variable (Quota) | -0.0632 | -0.0629 | 0.0003 |
| Sampling Frame Variable (Strata) | -0.0001 | -0.0001 | 0.0000 |
| Unobserved Variable (Neither) | -0.0014 | -0.0011 | 0.0003 |
| | | | |
| **Total Number of Schools** | | | |
| Contacted about participating | 187.3 | 187.8 | 0.5 |
| Excluded due to quotas | 160.4 | 160.8 | 0.4 |
| Invited to participate | 27.0 | 27.0 | 0.0 |
| Declined to participate | 27.0 | 27.0 | 0.0 |
| Agreed to participate | 0.0 | 0.0 | 0.0 |